# [1] Mechanics of magma chamber with the implication of the effect of CO2 fluxing

Simakin A.G.[1,2] and Ghassemi A.[3]

[1] Institute of Experimental Mineralogy, RAS, Chernogolovka, Russia, [2] Institute of Physics of the Earth, RAS, Moscow, Russia, [3] Reservoir Geomechanics & Seismicity Research Group The University of Oklahoma, USA

**Abstract:** As the magma ascends from its depth of generation to the surface, it is often stored in a series of chambers along the way. The rheological contrast between the viscous magma in the magmatic chambers and the surrounding rocks disturbed the stress field which can give rise to various modes of rock failure at magmatic pressures less than the lithostatic stress, leading to an eruption. Different modes of mechanical failure of the chamber walls are considered depending on the geometry and the sign the relative pressure. Relaxation of viscous stress around magmatic chambers, which is important on the time scale of weeks to months is considered in the analysis of stability with application to both large and extra-large magmatic chambers such as Yellowstone. The effects of a strong deep $CO_2$ flux in Yellowstone are considered in detail. The analysis shows that variations in the flow rate around the observed mean value of 40 kg/m2/yr in the hydrothermally active areas can change the composition of the magma for several hundred thousand years, and cause periodic uplift and subsidence of the caldera surface with a period of several decades.

Keywords: magma chamber, viscoelasticity, $CO_2$, eruption, Yellowstone

# 1. Introduction

## 1.1. Basic relations

In a first approximation a magmatic chamber can be regarded as the inclusion of a liquid in deformed elastic solid rocks. As a rule, viscous deviatoric stresses within this inclusion can be neglected and the stress state can be represented by hydrostatic pressure. The boundary between the enclosing rocks and magma can be approximately regarded as discrete. On this boundary the continuity of the normal and zero tangential stress conditions are valid (e.g., [1] ) or in vector notation:

$$\sigma \cdot n = P_m(t,x,y,z) \cdot n, \tag{1}$$

where n is normal to the chamber wall in point (x,y,z), $P_m(t,x,y,z)$ is the magma pressure. The pressure of the magma depends on the deformation of the walls of the chamber, and also on the mass and state of the magma inside it [1]:

$$M(t) = \int_{V(P(t))} \rho(P_m(t,x,y,z)) dV \tag{2}$$

Differentiating the r.h.s. and l.h.s. of eqn.(2) with respect to time while using the space averaged pressure $P_m(t)=P_0+dP_m(t)$ and expressing $V_{ch}=V_{ch,0}+dV(t)$, $M(t)=M_0+q_m t$ and $\rho=\rho_0(1+dP_m(t))$ yields another form of eqn. (2) which is used, e.g. in [2]:

$$\frac{dV_{ch}}{V_{ch,0}} = -\frac{dP_m(t)}{K_m} + q_m dt/V_{ch,0}\rho_0 \tag{3}$$

where $K_m$ is effective bulk modulus of the multi-phase magma, $V_{ch,0}$ and $\rho_0$ are reference magma chamber volume and magma density, $q_m$ is mass flux of magma. When magma contains fluid bubbles its compressibility increases significantly [2]. Change of the magma chamber volume at the hosting rocks deformation can be expressed in effective form used in theoretical analysis as equation of state (dependence of the volume from the magma pressure $P_m$, which is a reference pressure in the apical part of chamber boundary when hydrostatic pressure gradient is included)

$$dV_{ch}/V_{ch,0} = dP_m(t)/K_{ch} \tag{4}$$

A chamber volume at the rising pressure increases not only due to the rocks compression as suggested e.g., in [2] but also due to the uplift of the surface and roof buckling. Therefore effective chamber bulk modulus $K_{ch}$ can be at least order of magnitude smaller than that of the host rocks. Roof faulting reduces $K_{ch}$ several times more [1]. In the correct numerical models boundary condition (1) and EOS of magma (2 or 3) are applied simultaneously (fully coupled system of equations) and pressure is determined iteratively at each time step.

The mechanical modeling of the magma chambers is mainly aimed at:

1. Prediction of localization and mode of the failure of the host rocks including chamber roof, volcano cone or in more general term magmatic – tectonic coupling;
2. Localization of the transport of magma out and into the chamber by dykes;

3. Simulation of processes during a volcanic eruption or on the way to an eruption, including the mechanical effects of filling the chamber with melts and fluids, the magma outflow through dykes, the magma crystallization and the degassing.

**1.2. Tectonic-magmatic coupling**

A strong rheological contrast between the magma and the rocks leads to the localization of the strains around the magmatic chamber, so that the rock deformation (and failure) is coupled with the magma emplacement on a global scale, especially in an extensional regime [3]. In a compressional tectonic setting the magma chamber attracts thrusts, as shown in analog experiments and explained with an analytical solution for a circular hole in a compressed solid [4]. Simakin and Ghassemi [5] have studied numerically in 2D the effect of an elliptical magma chamber at shearing and showed that when interacting with the magmatic chamber, the strike-slip faults have experienced offset, similar to the step-over pattern.

In some situations the influence of a pressurized magma chamber on a regional stress field can be represented by the solution of the problem of a pressurized circular hole in an elastic plate subjected to uniaxial stress at infinity [6]. The calculated directions of the maximum compressive main stress (s1) are initially perpendicular to the chamber walls and, with an increasing distance from the chamber, rotate to the far-field stress direction. The dykes propagate through magma fracturing along the s1 directions and perpendicular to the least compressive main stress (s2 in 2D) [7]. Taking into account the influence of dykes on the stress field makes theoretical prediction of the dyke directions almost identical to those observed for the West Peak intrusion, Colorado [6].

**1.3. Influence of shape**

The mechanical properties of the magmatic chambers beneath the volcanoes depend on the shape of chamber, its relative depth (ratio depth to width), the gravitational load of the volcanic cone over the chamber. In general, magma chambers are horizontally elongated, as can be judged by the shape of solidified intrusions [8]. However, small shallow chambers under volcanoes often have an isometric shape close to a spherical. There are several theoretical studies in which the interaction of such spherical chambers with the surface is analyzed.

*1.3.1. Spherical and cylindrical chambers*

A closed-form analysis (improving classic Mogi [9] solution) of the stresses around a spherical inclusion (3D) in an elastic half-space was presented by McTigue [10]. The model predicts that the conic dykes (magma-fracturing) can be expected to initiate on a ring of maximum $\sigma_{\phi\phi}$ on the sphere's surface. The ring is delineated by planes (lines) drawn tangent to the sphere through a point on the symmetry axis at the surface. An accurate analytical solution for cylindrical chamber (circular in 2D) in the elastic half space was presented in [11]. Based on their solution for direction of the most compressive principle stress, authors developed conception of the capture zone for the ascending magma. They suggested that the magma ascending from the generation level in the zone of the differential stress above 1 MPa, caused by the presence of magma chamber, moves along s1 directions towards it. Overpressurized spherical magma chamber redistributes s1 directions in its vicinity normal to its surface and thus attracts ascending magma.

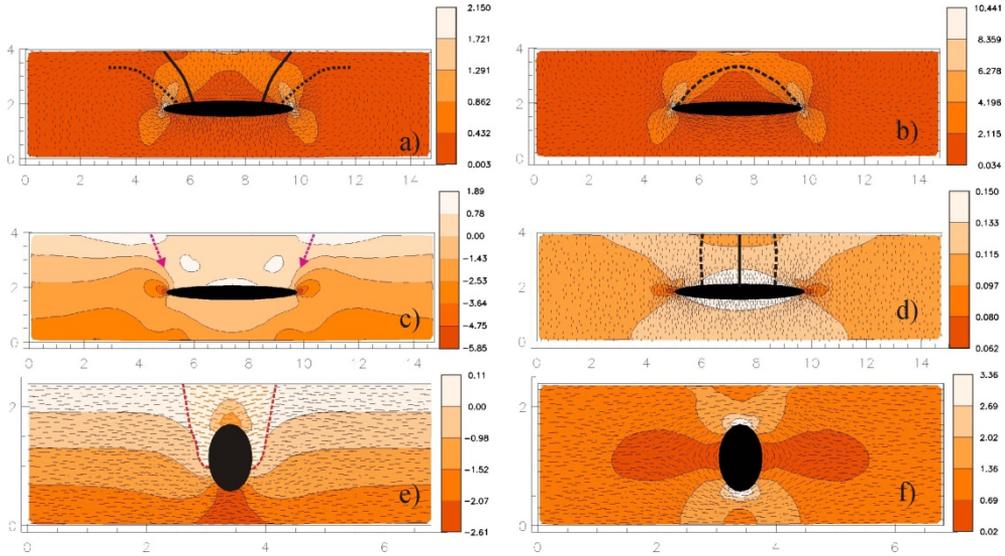

**Figure 1.** Examples of the calculated numerically in 2D elastic approximation of the stress fields around magma chambers without edifice. Spatial unit $l_0$ is arbitrarily chosen. Stress units are scaled with $\rho g l_0$, nondimensional Young modulus is 500 (at $l_0$=2500 m, $\rho$=2500 kg/m$^3$, E=30 GPa), Poisson's ratio ν is 0.2. Density of the magma and rocks are equal. Short dashes map s1 directions. a) s1-s2, overpressurized chamber potential dykes trajectories are localized near the roof edges b) s1-s2 in the undepressurized magma chamber, reference pressure in the apical point is zero, hydrostatic gradient exists c) s2 in the undepressurized magma chamber, red arrows depict normal faults starting on the surface d) s1-s2 at extension, black lines show potential trajectories of dykes out of magma chamber e) s2, overpressurized vertical chamber, potentially rocks within red dashed line can be ejected with diatreme formation f) s1-s2, underpressurized vertical chamber.

### 1.3.2. Elliptic chamber

The distribution of stresses around elliptical chambers near the surface has been numerically studied by many researchers (see in [12]). The main features of such solutions (in 2D) are shown in Figs. 1 and 2. Both the over-pressurized (Fig. 1a) and under-pressurized (Fig. 1b) elliptic chambers when interacting with a free surface create an arch pattern of the maximum differential stress (s1-s2). This parameter is twice larger than the maximum tensional principle deviatoric stress used in [11]). At the overpressure (Fig. 1a), the preferred paths of dykes start at the edges of the chamber. Only a small part of the s1 trajectories leads to the surface, while the other ones form the "saucer edges", known for shallow sills [13]. For the underpressurized chamber, the arch zone of the maximum shear stress (Fig. 1b) can cause the detachment of the apical roof zone into the chamber. On the map of s2 values (Fig. 1c), one can note zones of the tensile stress on the surface above the edges of the chamber. Many researchers [12, 14, 15] suggest an essential role in normal inward dipping ring faults in the roof collapse and caldera formation. When the regional fault crossed the roof of the chamber in a position parallel to the arch of maximum s1-s2, the stress is amplified and this configuration can initiate the trap-door caldera formation at an enough overpressure [1]. During the extension the conjugate normal faults tend to develop in the roof. The magma transport directions are vertical, and they all lead to the surface (see Fig. 1d). A vertical chamber with excess pressure creates a stress field (Fig. 1e), which can potentially lead to diatreme formation. The underpressurized vertical chamber only slightly disturbs the stress field (Fig. 1f).

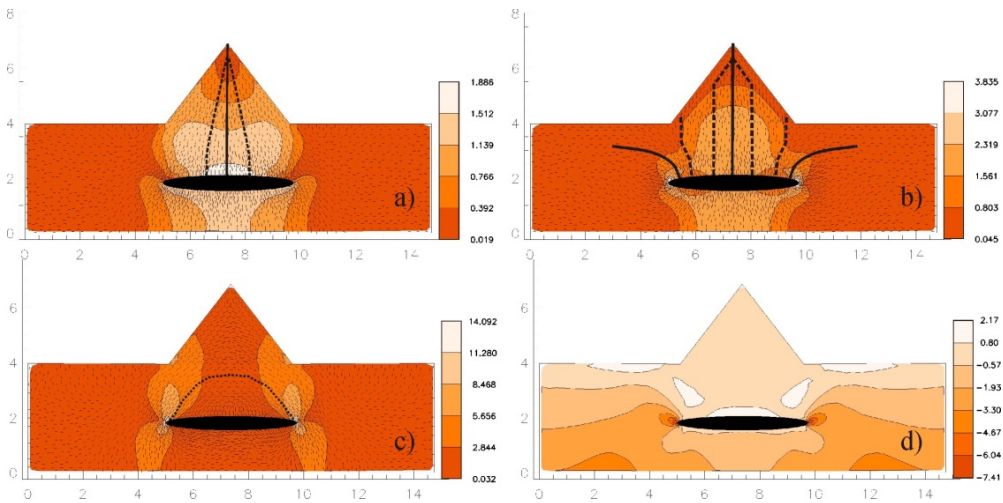

**Figure 2.** Calculated elastic stress field for magma chamber with edifice (calculations details as in Fig. 2) a) s1-s2 uncompensated and unrelaxed respond on the large edifice load, magma is focused towards cone top b) s1-s2 magma chamber is overpressurized, dykes trajectory towards the slope of cone appear c) s1-s2, reference $P_m$=0, edifice slope instability is expected d) s2, reference $P_m$=0, inward dipping normal faults formation is dictated by the extended linear zones with positive (tension) s2 in the roof.

*1.3.3. Elliptic chamber with cone*

A perfect volcanic cone is often formed with the repeated effusive eruptions from the magma chamber. The cone loading significantly modifies the stress field. Without overpressure, the cone focuses the trajectories of the magma towards the apex of the cone (see Fig. 2a). This theoretical prediction looks reasonable, since the huge perfect cones of Fujiyama and Klyuchevskoi volcanoes have a height of about 4 km above the base and were formed in hundreds of the repeated eruptions with magma outflow through the central crater at the top. However, at a strong overpressure, focusing is weakened, and some trajectories can lead to the slope and to the base of the volcano (see Fig. 2b). At the same time Pinel and Jaupart [16] predicted that the dykes propagating vertically are pushed out from the edifice edges, probably because they didn't include the magma chamber in their model. At the underpressure, the extensional zones develop on the surface and in the roof. Outward dipping zones of shear fractures also arise including the cone slope bottom (see Fig. 2c). The failure of the bases of the cone can lead to the instability of the slopes of the volcanic edifice and to the consequent emplacement of debris avalanches, followed by an explosive eruption and by the formation of a horseshoe-shaped caldera. In this case, another failure mode is represented by extensional fissures dipping inwards (Fig. 2d).

**1.4. Effect of viscous relaxation of the deviatoric stresses**

All the shown modes of the interactions of the magma chambers with the hosting rocks have been obtained in an elastic approach as an immediate reaction on the magma arrival or its evacuation. The long term deformations around the magma chambers should be treated with accounting for the stress relaxation. The viscoelastic Maxwell rheology is appropriate for the modeling of the rocks with temperature above c.a. 400°C at enough long times. The characteristic time scale of a viscous relaxation is a ratio of the viscosity to Young modulus of the heated rocks $\tau_{ve}=\eta/E$ that is 1-30 years at $\eta$ =1-10 $10^{18}$ Pas, E= 10-30 GPa. For example, geodetic data on the unrest in the in Long-Valley caldera in 1998 in the shorter time scale of several months were well interpreted accounting heated zones with viscosity $\eta$ =(1-100) $10^{16}$ Pa s [17].

In the two-dimensional numerical model by Simakin and Ghassemi [1] the viscosity of the rocks was calculated for the quasi-steady state temperature field around an elliptic magma chamber. It was shown that at the constant influx rate (without magma loss into the dykes and at the eruptions) the magmatic overpressure (and corresponding deviatoric stresses in the hosting rocks) soon approaches an asymptotic value depending on the current magma chamber volume. This model is incomplete, since the paths leading to the probed current state of a chamber of a given size are not taken into account. Karlstrom et al. [18] have included the transient temperature field and the dyking in their complex model of magma chamber evolution. The mechanics was modeled with a rather complicated analytical solution for the circular inclusion in the viscoelastic half-space [18]. Obviously many provisional features of viscoelastic shell were introduced due to the formal symmetry requirements. The model is not fully consistent, since the undefined initial chamber size has been used as a parameter in the evolutionary model. Melting and assimilation description is crude in comparison with special studies. Viscosity used in the calculations is not explicitly shown. Nevertheless by the order of magnitude results of both studies are quite close. According to [18] flux of about 0.004 km$^3$/y is required for a chamber with volume 37 km$^3$ to initiate eruption, similar parameters are obtained for the reference viscosity of the rocks in thermal aureole of $\eta$=5 $10^{19}$ Pa s at T=400°C [1]. This value is within a range of the typical magma supply rates for continental silicic magmatic centers of 4.4±0.8e-03 km$^3$/yr [19].

Both the models predict that for a large magma chamber the magma fluxes as high as 0.1-1 km3/a are required to reach an overpressure capable to cause roof failures and caldera-forming eruptions. In this chapter we will consider additional factors that can influence the thermal and mechanical state of large magmatic chambers.

## 2. Deep fluid flux as an important mechanical factor

There is an increased interest in the giant silicic magma chambers, which in their life can become a batholith or a supervolcano, or both. Since in historical time there were no supereruptions with a volume of 1000 cubic kilometres and more, judgments about the mechanisms of this phenomenon remain essentially theoretical. However, some aspects of this problem are almost completely clear. A huge volume of silicic magma can be formed by melting a silicic crust at the underplating with basaltic magma, which may include remelting of former rhyolitic volcanic rocks and granites. For the volcanic complex Altiplano-Puna of the central Andes, including the caldera of La Pacana, with size 65x35 km, this mechanism is convincingly shown by geologic and geochemical data [20]. The Altiplano-Puna complex was formed at a thickening of the continental crust and intensive basalts generation in the mantle wedge at the steepening of the subduction angle and probably delamination of the crust [21]. For Bishop Tuff rhyolites (Long Valley, California), the composition of radiogenic isotope indicates a significant mantle contribution, which is interpreted as the consequence of the midcrustal melting of amphibolites with basaltic protolith of mantle origin [22]. Another possible explanation for these observations may be the contamination of the silicic melts by REE, Pb and Sr, transferred by the $CO_2$-CO fluid from the underplating basalts. Below we will consider some of the already established geochemical consequences of the fluxing of rhyolitic magma with deep $CO_2$. In any case, a sufficiently high rate of basalts generation is an essential condition for the production of a large volume of rhyolites. This requirement can be fulfilled in deep mantle plumes or in smaller upper mantle plumes (ascending flows) caused by the crustal delamination or breakage off the oceanic plate in the collision zones. The mechanics of assembling large volumes of silicic magma from small portions is analysed in [23, 24]. In a long term, magma in giant chambers is in a state of mush.

### 2.1. Why mush

#### 2.1.1. Theoretical view

The melt in the partially molten zones of the Earth tends to separate from the solid phase when their densities are different. A two-phase melt-crystals homogeneous system is gravitationally unstable. At low concentration of the solid phase, the crystals separated from the magma by Stokes settling. When the solid phase content becomes close to 40-45 vol.%, the crystals begin to interact mechanically and form a coherent matrix. At the melt density lower than that of matrix, melt separation takes place in the form of a compaction. The physics of compaction is quite complex and deserves special consideration, which is beyond the scope of this chapter. For the subjects considered below, it is important to know the stable content of the melt in the mushes of basic and silicic compositions. The effectiveness of the compaction is characterized by the compaction length $\delta_c$ [25]:

$$\delta_c = \left[\frac{\varsigma + \frac{4}{3}\eta_s}{\eta_m} k(\varepsilon)\right]^{1/2} \qquad (6)$$

where ζ and $η_s$ are the bulk and shear viscosity of the matrix, respectively, $η_m$ is magma viscosity, k is a permeability of the matrix with porosity ε. When the thickness of the mush layer is less than $δ_c$ the compaction is fast. In a detailed analysis of the matrix viscosity and permeability Tegner et al. [26] estimated the compaction length for olivine-plagioclase cumulates of Skaergaard intrusion at an initial porosity of 22 vol.% at $δ_c$ =870 m. Due to the fast compaction in the geologic sense with the mush thickness < $δ_c$, the volume fraction of the residual melt in the olivine-plagioclase orthocumulates of the Skergaard reduced to several volume percent [26]. This estimate is consistent with McKenzie's [25] conclusion that in the regions of basalt generation in plumes and in Mid-Ocean Ridges, the volume contents of the melt is no more than a few percent due to the rapid compaction and segregation of the melt into the veins and dykes. In silicic systems, compaction is less effective because of the low viscosity of a wet mush with a high quartz content and relatively high viscosity of the rhyolite melt. The corresponding meter-decimeter scale of $δ_c$ and nonlinear rhyological effects lead to local phase separation with the formation of a migmatite like texture [27]. It is usually believed that additional tectonic deformations are involved in the separation of granitic magma and the formation of displaced granite plutons [28]. Thus, while in the basaltic system, partial melts with a volume fraction of melt greater than a few percent exist only temporarily [25], silicic mush, probably locally heterogeneous, with melt content of 10-20 vol.%, can persist for long time, limited by the rate of cooling and solidification.

The erupted Yellowstone intracaldera rhyolites are fairly homogeneous [23]. Periodic intrusion at different levels in the mush zone of superheated rhyolites from contact with underplating basalts can be a physical mechanism that wipes the mush heterogeneities and provides mixing on a large scale. The direct invasion of basaltic magma into the mush was hardly massive, since in Yellowstone only traces of clots of mafic minerals were identified in erupted intracaldera rhyolites [29], in contrast to the mixing textures well expressed in the extracaldera rhyolites [30].

*2.1.2. Geophysical observations*

A geophysical insight into the state of large silicic magma chambers requires a sufficiently large volume of observations. The seismic tomographic model of the Yellowstone volcanic system with two magmatic columns under resurgent domes merging at depth was proposed by Husen et al. [31]. It was significantly improved when data were combined from the dense seismic arrays of the Yellowstone, Teton, and Snake River Plain (SRP) regional seismic networks, the NOISY array, and the wide-aperture EarthScope Transportable Array [32 and refs. in it]. A continuous layer of a partial melt under the caldera at depths of about 6-12 km was clearly identified with a separate, presumably basaltic magma accumulation zone at depths 20-50 km. In this study, the volume fraction of the melt in the upper crustal body was estimated at 9 %, and in the lower zone - 2 vol.%. Another independent source of information is provided by the magnetotelluric method. For the Yellowstone upper crust, the highly conductive zone of the hydrous partial melt is localized at approximately the same depths 6-12 km under the caldera [33, 34], as envisaged by seismic tomography. In fact, the interpretation of MT data is not absolutely unambiguous. High conductivity can be interpreted as a manifestation of a partial silicate melt or carbonatitic melt [35], aqueous or carbonic fluids, graphite. MT and seismic data may be complementary since the same volume fractions of carbonatitic and silicate melts will have a similar effect on the seismic velocity, while the conductivity of carbonatite melt will be at least two orders of magnitude higher [35]. Therefore, a very small volume fraction of carbonatites (or $CO_2$-CO based solutions) will be detectable by MT and not visible in seismic data. The observed low resistance layer (seismically practically undistinguishable) under all SRP to the west of the Yellowstone caldera at depths of about 50 – 70 km can be a residual carbonatite melt with a low melting temperature or concentrated solution in carbonic fluid, rather than a basaltic partial melt.

In this chapter, we will focused on analyzing the large magma chambers created in some way such as the chamber under Yellowstone volcanic field which is one of the most studied (see e.g., review [36]). A generalized scheme of such a silicic chamber formed above the hot spot is shown in Fig. 3. The possible role of volatiles in the heat budget and the mechanics of large silicic magma chambers is the least studied and will become the main subject of the following paragraphs.

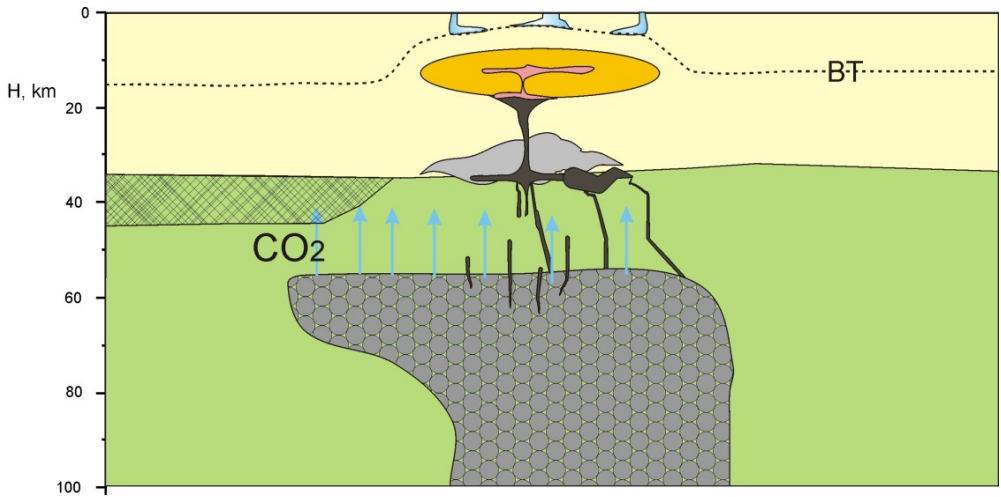

**Figure 3.** Cartoon schematically displays large silicious magma chamber in the continental hotspot setting. Basaltic magma originated in plume underplates crust (black) and where it partially solidifies to form mafic intrusions and partially rise to mush zone. At the lower contact of the mush zone superheated rhyolite is formed and invades into mush. Hydrothermal system operates in the upper crust above brittle- ductile transition level. $CO_2$ infiltrates through the zone of magmatic activity and accumulates near the Moho away of this zone (crosshatched).

## 2.2. High emission of deep CO2

$CO_2$ is the main component that is released during degassing of basalts, especially from dry basalt formed during decompression melting in plumes. It was also assumed that in the lower crust there are reservoirs of the supercritical $CO_2$ derived in the mantle at 2-3 GPa [37]. A direct observation of the $CO_2$ mantle reservoir (at c.a. P=1.8 GPa) sampled in the xenolithes from paleovolcanoes in the Panonian basin is reported in [38]. In these xenoliths $CO_2$/magmatic glass mass ratio is estimated of up to 0.25. The carbonated silicate melt with such $CO_2$ a content releases $CO_2$ at P=2-3 GPa and later at their ascend the basaltic magma and the $CO_2$ can move independently. Where deep sources of $CO_2$ exist, the magma chambers will be subjected to high fluxes of volatiles, which can play a significant role in keeping long-term heat balance [39], in modifying the composition of magma and in influencing the deformation of the magma chamber. Bachmann and Bergantz [40] attempted to model the essentially water fluid Darcy's flow through the partial rhyolitic melt. In the model water fluid was released from the basalt magma with 4-6 wt.% of dissolved $H_2O$. They found that possible fluid contribution in the instant mobilization of several thousand $km^3$ of mushes immediately before eruption is negligible. Their conclusion is mainly determined by slow Darcy's filtration rate while faster mechanisms can exist.

In the following we will consider effect of $CO_2$ fluxing on the example of Yellowstone caldera since the total $CO_2$ flux through Yellowstone caldera is among the largest and is about an order of magnitude greater than across the entire rift zone of Iceland [41]. Werner and Brantley [42] found that the largest flux of 456/kg/$m^2$/yr in average is registered in the areas of the acid geotherms with surface 145 $km^2$. Average $CO_2$ flux in the hydrothermal activity zones is about 40 kg/$m^2$/yr while in inactive forestry area background flux is 3.5 kg/$m^2$/yr. Authors ascribe isotope composition of carbon and helium to 70% of $CO_2$ from the mantle source and the rest from sediments (Mamoth limestones). An analysis of $CO_2$ flux value showed that it is significantly larger than expected at the basalts magma degassing one at the rate of basalts generation in plume 0.03-0.05 $km^3$/yr [42]. Extraordinary $CO_2$ generation can be linked with interaction of the mantle plume and Farallon pate. Part of subducted slab slides along 400-600 km phase boundary under SRP [43]. It is separated by lateral tear from the older part of the slab intersecting this boundary. Plume crosses Farallon plate trough the tear. Presumably oceanic slab contains carbonates that are mobilized by hot plume material and release $CO_2$ at pressure below 2-3 GPa (see e.g., in [44]).

**2.3. Heat budget with CO2**

The heat supply by deep fluids may be important in order to prevent from cooling the large volumes of the partial melts in the large and super-large magma chambers (such as the Yellowstone In the most general form the heat budget of the mush layer can be estimated adding the heat supplied by the different contributions of the carbonic fluid and basaltic magmas. The dissipation of heat depending on the specific physical mechanism can reduce the positive part of the balance. The heat lost is assumed to be essentially vertical, equal to the observed average heat flux, while the horizontal thermal loss is neglected. At the positive heat balance the silicic melt will be generated and probably erupted, while at the negative balance the magma will solidify approaching a scenario of formation of batholiths.

The amount of the advective heat transfer by the ascending fluid depends on the mechanism of transport. If the fluid has passed the interval from contact with the basaltic magma to the layer of the partial rhyolitic melt along the fractures rather than by the Darcy's flow, then the heat losses through the crack walls will be small, and we can assume that the heat flux is approximately equal to

$$Q_{adv} = c_{p,fl}(1200-850)q_{co2} = 0.22e08 \text{ J/m}^2/\text{yr } (0.70 \text{ W/m}^2), \quad (7)$$

where $q_{co2}$ is mass flux of $CO_2$ in kg/m²/yr, the specific heat of the fluid $c_{p,fl}$ is taken equal to 1.4 Kj/kg/o, the temperatures of basaltic magma and rhyolitic mush are 1200 and 850°C, respectively. As the first proxy, $q_{co2}$ = 40 kg/m²/yr is taken equal to the average value of flux observed in the geothermal regions of Yellowstone [42]. Below we consider the effect of the variation in $q_{co2}$ over a wide range on the heat balance.

In fact, the carbonic fluid exsolving from the underplated basalts and stored in the mantle reservoirs is a reduced. At P=2-3 kbar, the mole fraction of CO in such a fluid at $f_{o2}$ around QFM-0.5 is in the range 0.1-0.2 depending on the temperature [45]. At the oxidation of CO heat is released:

$$CO + 1/2 O_2 = CO_2 \quad (8)$$

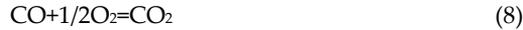

The enthalpy $\Delta H_T$ of the oxidation reaction (8) at T=700°C is -198870.4 J/mol. Volume effect of the reaction (8) is negative, so the heat effect is somewhat larger than $\Delta H_T$ at a pressure P= 2 Kbar corresponding to a chamber depth of 7 km. However, not pure oxygen can be involved in the reaction, so we use the value of $\Delta H_T$ as the maximum estimate. If we take $q_{co2}$ equal to 40 kg/m²/yr, as above, and set $X_{CO}$ = 0.1, then the heat of CO oxidation will be equal to

$$Q_{CO} \approx -0.1 \, \Delta H \, q_{co2} 1000/44 = 0.18e08 \text{ J/m}^2/\text{yr } (0.57 \text{ W/m}^2) \quad (9)$$

The main source of heat in Yellowstone is associated with the flow of basaltic magma. The simple arrival of $q_b$ = 0.05 km³ of magma annually means advective heat flux

$$Q_b = q_b(\rho_b c_{p,b}(1200 - 850) + \Delta H_{m,b})/S_0 = 0.51e08 \text{ J/m}^2/\text{yr } (1.62 \text{ W/}m^2) \quad (10)$$

where the surface of the active part of the caldera is roughly estimated at 2000 km², the density of basaltic magma is set 2700 kg/m³, the specific heat $c_{p,b}$ is 1 kJ/kg/o and the latent heat of fusion, absorbed at melting of rhyolites is $\Delta H_{m,b}$=400 kJ/kg. With these values of the parameters, the value of the advective heat flux $Q_b$ is comparable in order of magnitude with the heat flux associated with $CO_2$.

Basalts generally don't intersect partial rhyolite layer serving a density filter. Their heat content is transferred to the superheated rhyolites formed on the lower boundary of mush zone that can penetrate in the partial melt. While fluid is not accumulated in the crust and rise through partial melt transferring heat and dissolved components into and out of the mush zone. Purely conductive heat transfer through partial melt and upper thermal aureole of magma chamber is limited by the stationary heat flux through mush layer in the absence of convection of the order

$$Q_{cond} = \lambda \Delta T / H_{mush} \quad (11)$$

where $\lambda$=2.5 W/m/o, $H_{mush}$= 6 km. With these parameters values, the conductive heat flux will be only 0.145 W/m² or 0.05e08 J/m²/yr. The average heat flux through Yellowstone caldera is significantly larger than this conductive estimate.

The dense basaltic magma interacting with the lower contact of the mush zone transfers its heat content to the superheated rhyolites formed during this interaction. The superheated rhyolite penetrates into the mush zone and indirectly transfer $Q_b$. The low density fluid moves directly through the lower contact into a partial melt and transfers heat and dissolved components.

## 2.4. Thermal profile through the mush zone

The above global estimates are valid regardless of the real physical mechanism of fluid transport. A self-consistent mechanical model of fluid transport through a partial melt should include the coupled magma and fluid flow through a deformable viscoelastic matrix that extends the viscous [25] and visco-pastic [27] compaction models. With a sufficiently comprehensive model one can consider complex flow regimes similar to a self-organized critical transport observed experimentally [46]. Below, we consider the simplest estimate of the spatial temperature distribution in a mush with a pervasive fluid flow as a solution of the general advection problem in a porous media. The mass flux of the fluid will not be found as a part of the solution (e.g., as a solution of Darcy's equation, as in [40]) but is specified on the basis of observations. In the general form, 1-D equation of advective heat transfer through a porous medium is

$$\rho_s c_{ps} \frac{\partial}{\partial t} T(z,t) = \lambda \frac{\partial^2 T(z,t)}{\partial z^2} + c_{pf} Q_f \frac{\partial}{\partial z} T(z,t) \quad (12)$$

or in nondimesional form

$$\frac{\partial}{\partial t} \bar{T}(z,t) = \frac{\partial^2 \bar{T}(z,t)}{\partial z^2} + Pe \frac{\partial}{\partial z} \bar{T}(z,t) \quad (13)$$

where Peclet number is $Pe = l_0 c_{pf} Q_f / k_T \rho_s c_{ps}$. Substituting $l_0$=200 m, $k_T$=1.0e-06 m²/s, $c_{pf}$=1.4 kJ/kg/o, $c_{ps}$=0.84 kJ/kg/o, $\rho_s$=2500 kg/m³ we get the time scale 1300 yrs, Pe=0.013-0.4 with $q_f$ in the range 4-100 kg/m²/yr. At this time and spatial scales problem is weakly to moderately advective (Pe<1). It is important to set proper boundary conditions for temperature.

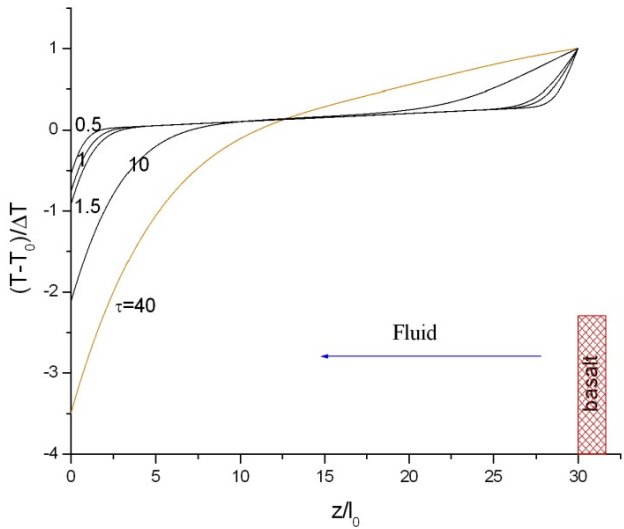

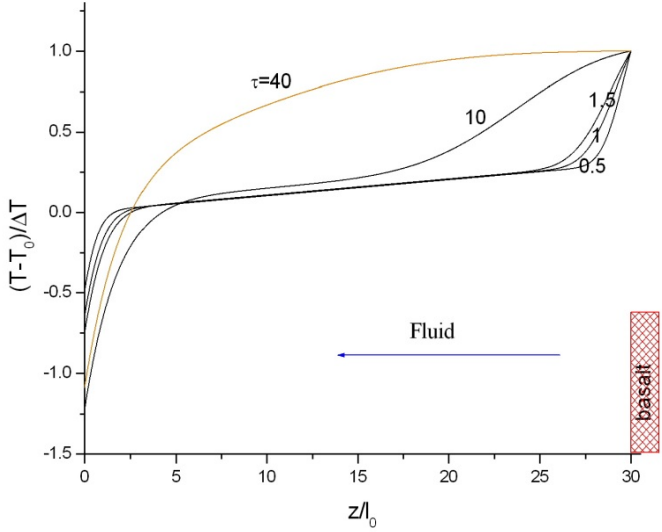

**Figure 4.** Calculated temperature profiles in the mush zone in nondimentional form in the moments of time τ=0.5, 1.0, 1.5,10, 40. Length scale scale $l_0$=200 m, time scale $t_0$=1300 years. Right boundary is at T=$T_b$=1200°C, left boundary is subjected to the constant heat flux of 2.3W, initial temperature of mush $T_0$=850°C a) Pe=0.13 (corresponds to $q_{CO_2}$=40 kg/m²/y b) Pe=0.5. In a) at τ=40 temperature at left boundary is physically meaningless, since constant heat flux boundary condition does not guarantee even positive absolute temperature. At Pe=0.5 left boundary temperature stabilizes around reasonable T=500°C, however then almost one third of the mush zone asymptotically is heated to $T_b$.

At the lower boundary, the temperature of basaltic magma $T_b$=1200°C is prescribed. The heat flux from the mush layer supports the heat flux in the geothermal system and in the quasi-steady state their values should be close in average by area and time (as used in 1D models). In our calculations, a heat flux $Q_h$ of 4.6 GW/2000km² =2.3 W/m² at the upper boundary is set, equal to the average estimates for Yellowstone [36].

We calculated the transient solutions for different Pe values. At Pe=0.13, corresponding to a mass flux of $CO_2$ of 40 kg/m²/yr. in a geologically short time of 13 Kyrs, the temperature at the top of the mush layer drops to 200°C and continues to decrease and the thickness of the mush layer is halved (see Fig. 4a). With a stronger fluid flow Pe=0.5 in time equal to 56 Kyrs, the boundary temperature still increased to c.a. 500°C, but did not reach a steady-state value (see Fig. 4b).

In our formulation, in steady state, there is relationship between Pe($q_{fl}$) and the maximum admissible heat flux $Q_{h,max}$. This is determined by the fact that temperature at the outer boundary (closer to surface) is part of solution. We fix a reasonable value of T(0)=-0.7 (about 600°C in physical dimension) and get

$$Q_h = 1.7 Pe / (1 - \exp(-30.0 Pe)) \tag{14}$$

The numerical solution agrees with equation (14), at Pe=0.13 the maximum $Q_{h,max}=0.226$ (eqn.14) is less than 0.73, used in the numerical solution, and at Pe=0.5 $Q_{h,max}=0.85 > 0.73$. As expected the mush layer solidifies in the first and melts in the second case, respectively.

The heat flux is measured directly only on a small part of caldera. The measurements of the heat flow at the bottom of Yellowstone Lake [47] gave values 100-300 mW/m² near the caldera boundary and up to 1600 mW/m² in Mary Bay. Fournier [48] obtained an estimate of 2.65 W/m² (with an active surface 2000 km²) on the basis of the correlation of the chlorine content and the enthalpy of the geothermal fluid. A later inventory of the chlorine balance in the Yellowstone river system [49] gives an average heat flux of 2.3-3.3 W/m² (when recalculated to an active caldera area S=2000 km²). The $CO_2$ flux was measured in geothermal areas at a quiet stage of the caldera evolution. The amount of $CO_2$ accumulated and periodically released is unknown. In fact, deep $CO_2$ is partially redistributed horizontally, accumulated underground and transferred as an $HCO_3^-$ ion by geothermal waters. The rate of basaltic magma supply ($q_b$) is also known with low accuracy. In view of all these uncertainties, we are considering the heat budget in a wide range of variation of the parameters involved and display the results in Fig. 5.

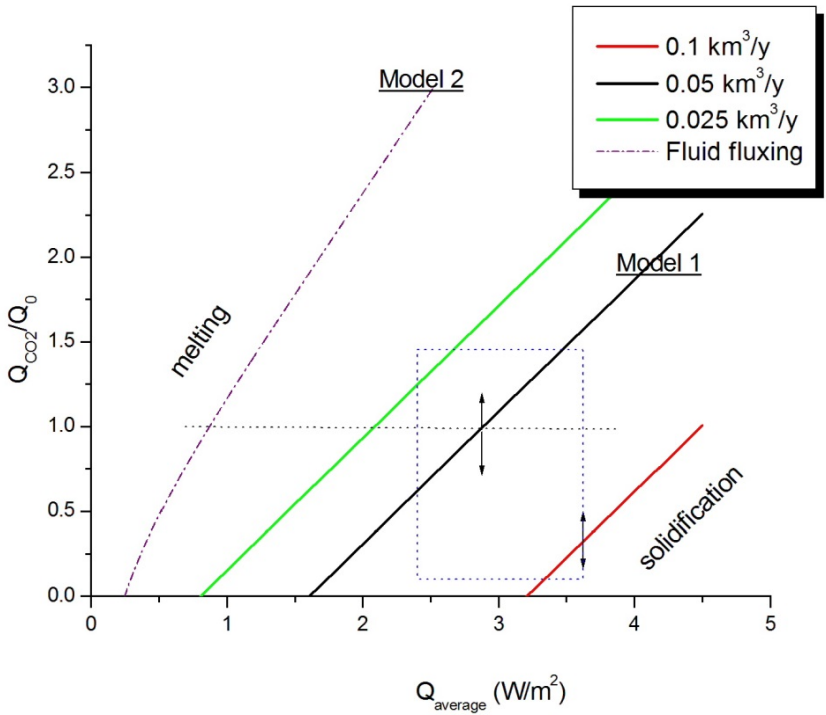

**Figure 5.** Zero heat balance condition for rhyolitic mush zone of Yellowstone volcano in the coordinates average surface heat and $CO_2$ fluxes a) continuous lines correspond to the different basalt magma supply rates and mechanisms providing full heat transfer of basalt magma and deep $CO_2$ fluid to the mush b) dash-dot line plots relationship for advective fluid heat transfer through the mush and passive

basalt underplating setting low boundary constant temperature T=$T_b$ condition. $CO_2$ flux is normalized on the $Q_0$=40 kg/m²/yr. At the heat flux larger than zero-balance value mush is solidified at the smaller heat flux or larger $CO_2$ flux mush is melted.

The zero heat balance conditions in the mush zone for different $q_b$ are indicated by solid lines in the coordinates $q_{CO2}$ - $Q_h$ (surface heat flux). In general, the system is close to the global balance at $q_b$=0.05 km³/yr. Doubling of this rate shifts it to the rhyolite magma generation. Variations in the flux of $CO_2$ can lead to the transitions from solidification to melting in a state close to thermal equilibrium. The dashed line shows the dependence of the heat and $CO_2$ fluxes calculated using eqn. (14). Evidently, when basalt is passively underplating the mush zone, providing T=$T_b$ (boundary condition in the advection model) only an extremely high $CO_2$ flux can ensure the long-term existence of the mush layer. This analysis emphasizes the importance of the interaction of mush with superheated dry rhyolites (see Fig. 3). In fact, convective melting and mixing is fairly fast process [50], so that zircon crystals from melted mush will survive. Mineral thermometers will record crystallization temperatures at a solids content of about 40 vol.%, when convective melting stops, well below the initial temperature of the superhested rhyolites.

## 2.5. Volatiles in the melt at $CO_2$ fluxing

Our analysis demonstrates that the flux of $CO_2$ in a reasonable range can make a significant contribution to the heat budget of the mush layer. When crossing a layer of mush, the fluid should create petrologic signs of interaction with hydrous rhyolitic magma, e.g., by transferring water from the melt to the $CO_2$ bubbles. During ascent and eruption, volatiles escape from the bulk melt, information about the pre and sin-eruptive conditions is retained in the melt inclusions in magmatic minerals. In the equilibrium with the fluid, the contents of $H_2O$ and $CO_2$ in the melt depend on the PT conditions and the composition of the fluid. We approximate the mutual solubility of $CO_2$ and $H_2O$ in the rhyolitic melt, using the compilation of experimental data from [51]. The solubilities in the melt are expressed through the pressure (in Kbar) and the mole fraction of $CO_2$ (z) in the fluid as:

$$C_{H2O} = 4.1\sqrt{P(1-z)^{1.35}}, \quad C_{CO2} = 0.0499 k P z^{0.75}, \quad k = -0.54013 + 0.0164T, \qquad (15)$$

where C is in wt.%, temperature - in °C. The widely used model of the mutual $CO_2$-$H_2O$ solubility [52] in the rhyolitic melt is more accurate, but also more complicated and can not be incorporated in numerical codes, such as a two-phase magma flow simulator. Our short expressions are for these numerical applications.

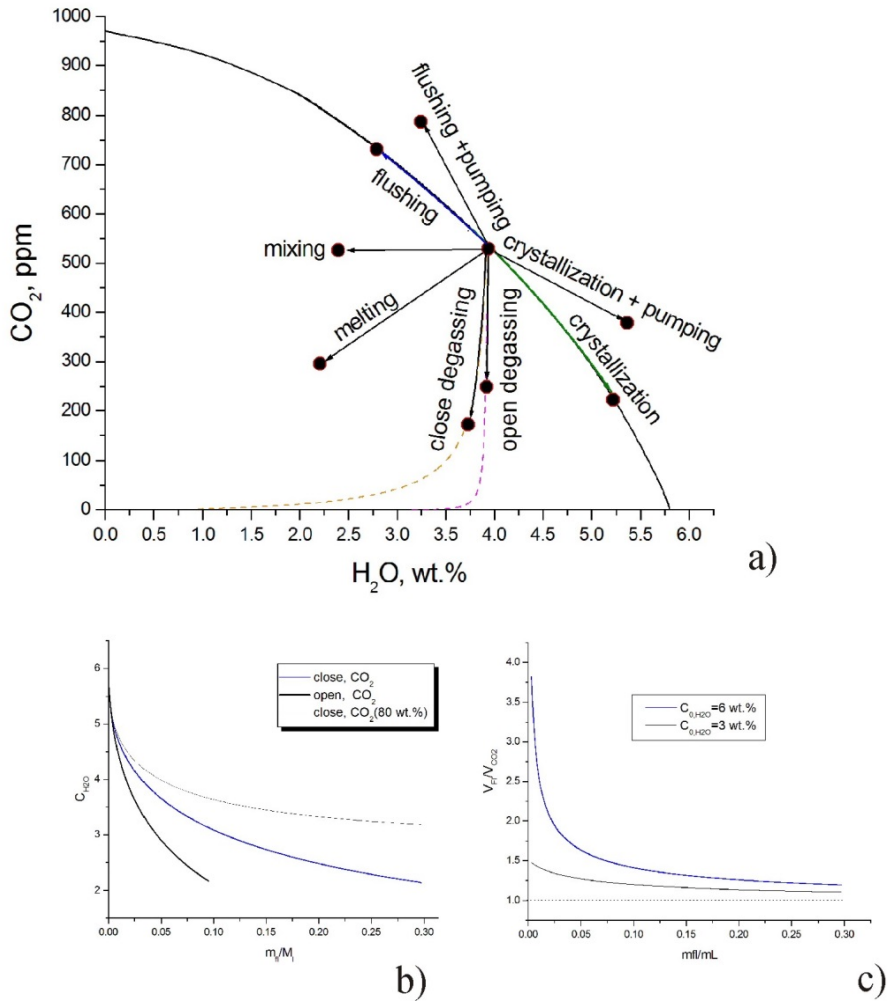

**Figure 6.** Results of modelling a) different factors affecting volatiles content in the melt b) dependence of the water content in the magma at the flushing with $CO_2$ on the mass ratio of the reacted $CO_2$ and melt. Style of the interaction and content of $CO_2$ in the incoming fluid are shown in the legend c) ratio of the volume of $CO_2$-$H_2O$ composition fluid to the initial volume of $CO_2$ as function of the mass ratio of the reacted $CO_2$ and melt. In the legend initial water contents in the melt are shown.

To model the melt and fluid equilibria, we solve the equations of mass balances for $CO_2$ and $H_2O$ along with Eqs.(15). In the Fig. 6a the isobar for P=2 Kbar for the composition of the rhyolite melt in coordinates $C_{H2O}$ – $C_{CO2}$ is plotted, the maximum concentration of water and $CO_2$ are for the pure water and $CO_2$ fluid, respectively. Various processes affecting volatiles in magma have been modelled and are depicted in Fig.6a. When the melt is depressurized during the eruption, degassing begins with the shift of the melt composition along the trajectories of degassing under the open (when all exsolved fluid is removed) or close system. Trajectories are shaped by the early loss of $CO_2$, followed by the release of $H_2O$ (more on this in [52]). Melting of the mush as a close system leads to dilution of the melt with respect volatiles (movement of the composition from the initial point to the origin of coordinates in Fig.6a). Crystallization at constant pressure leads to a shift to more water-rich compositions along the isobar. If the fluid exsolved during crystallization remains in the mush, the pressure rises. The result of mixing with superheated rhyolite depends on the composition of the later. A transition to water poor, fluid undersaturated compositions is expected in general. At $CO_2$ fluxing at constant pressure, the composition follows isobar from the water-rich to water poor melt. During CO2 flushing, the pressure of the magma can increase due to the expansion of the chamber being constrained by the host rocks. Then the trajectory of the melt composition in H2O-CO2 space will deviate from the isobar (see Figure 6a). A significant vertical size of the magma chamber can also cause deviations of the MIs compositions from the isobar [53].

The parameters of the $CO_2$ flushing process require more detailed consideration, since they are necessary for further mechanical analysis. The modeling of the closed system flushing was carried out by increasing of the total contents of $CO_2$ and $H_2O$ in the system in small increments in the proportion of the fluid composition. To evaluate the volume of a fluid in an equilibrium two-phase system, we use EOS for a $CO_2$-$H_2O$ mixture from [54]. In Fig. 6b, one can see that added mass of the fluid increases, the concentration of water in the melt gradually drops. Pure $CO_2$ eventually will extract all the water from the melt. With a starting content of 6 wt.%, it is sufficient to add about 30% of $CO_2$ by mass to reduce the water content to 2 wt.%. In the open system approximation, after every step, all the fluid is removed, and a new portion of carbonic fluid reacts with the pure melt. In this case, approximately 10 wt.% of $CO_2$ is sufficient to achieve the same effect. When the incoming fluid contains 20 wt.% of $H_2O$, it is equilibrated with the melt with c.a. 3 wt.% of $H_2O$ and can not reduce the water content below this value. This implies that in the SRP system, carbonic fluid penetrating the rhyolite mush contains less than 20 wt.% of water to produce melts with a lower content. The transfer of water from the melt to the fluid changes the composition of the fluid and enlarges its volume. In Fig. 6c, the volume ratio of the final fluid and the incoming pure $CO_2$ at P=2 kbar is plotted with the two initial water contents. A strong volume increase of 1.5-4 times has place at the lowest mass ratio of the added fluid to the melt. As the mass ratio increases, the volume expansion becomes less significant. Extraction of water and CO2 dissolution in the melt at the constant temperature induce crystallization due to an increase in the liquidus temperature. Crystallization caused by water loss will release heat of fusion and heat the melt, so that the integral effect should be more precisely evaluated with numerical modeling, which is beyond the scope of this chapter.

### 2.5.1. **Melt inclusions data supporting the concept of CO2 flushed mush**

Fig. 7 shows the results of the studies of melt inclusions in quartz phenocrysts from various large-volume rhyolites from Snake River Plain (SRP) and Long Valley calderas (Western USA). Lava Creek Tuff (LCT, Yellowstone 0.64 Ma) [55] decompression sequence is rooted in a 2 kbar isobar with an initial relatively high water content of 3 wt.%. The post caldera Tuffs of Bluff Point (Yellowstone, TBP 175-180 kyr) [56] sequence corresponds to degassing at the decompression from a pressure slightly higher than 1.5 kbar with a starting water content of 2 wt.%. Early Arbon Valley Tuff (AVT) [57] represents the products of the Picabo caldera eruption (10.44 Ma) with the highest water content of 6.5 wt.%. The AVT-LCT-TBP sequence illustrates a gradual decrease in water and an increase in the $CO_2$ content. Bishop Tuff (BT) MIs [53] demonstrate a highly variable set starting in $CO_2$ free compositions with c.a. 6 wt.% of $H_2O$. Some points follow approximately an isobar of 2 kbar, while others lie on the open system degassing path. One exceptional point with the highest $CO_2$ content reaches an equilibration pressure of more than 2.5 kbar. In general, the BT data are consistent with the fact that the magma was flushed with $CO_2$ followed by an eruption. Alternatively, the observed $CO_2$ enrichment and $H_2O$ loss process can be explained by mixing superheated rhyolites with a high $CO_2$ content and hydrous partial melts (see Fig.6a). It is likely that AVT data with a simple shift of compositional points along the $H_2O$ axis may correspond to such a mixing process that creates a fluid undersaturated melt. For comparison, we plotted in Fig.7 the MIs in quartz data for the Toba volcano [58] located in the subduction zone. Obviously, in a subduction setting with much lower $CO_2$ fluxes and a shorter residence time, the magma contains much less $CO_2$. However, even in this case, one can expect a $CO_2$ fluxing with the alignment of some compositional points along the 1.5 kbar isobar. Other points may be located on the close system degassing trajectories, starting with an isobar of 1.5 kbar (see in the inset in Fig. 7).

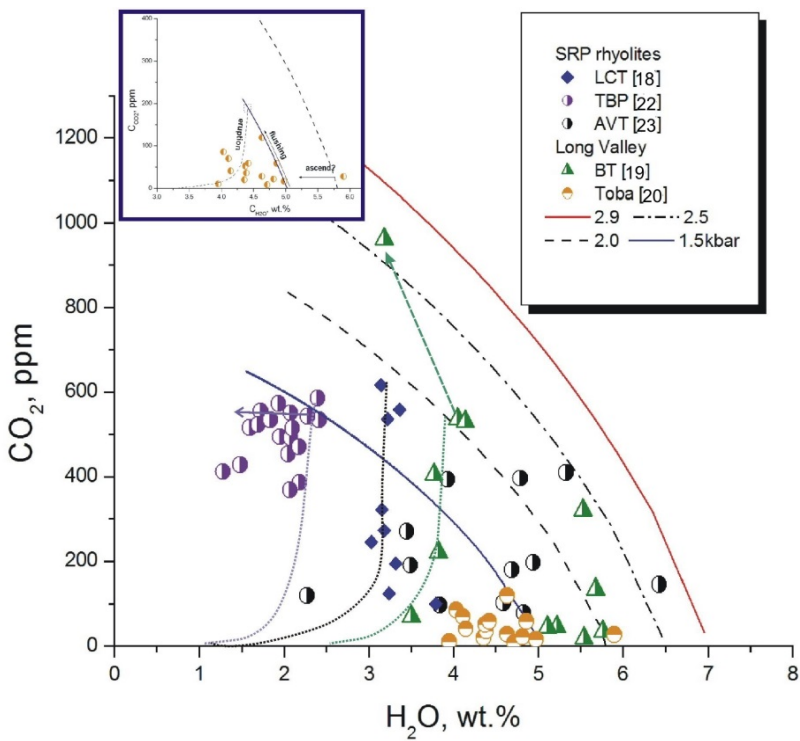

**Figure 7.** Contents of CO$_2$ and H$_2$O in MIs in quartz from silicic volcanic rocks from western USA and Toba. Line show equilibrium mutual solubilities of CO$_2$ and H$_2$O at the different pressures. Points located along isobars may correspond to the CO$_2$ fluxing of the hydrous silicic magma. Pyroclastics from the Bishop Tuff eruption (Long Valley caldera) experienced more strong interaction with CO$_2$ than Yellowstone system. In the Toba volcano localized in the subduction system this interaciton is weakest.

## 3. The "heavy breath" of the Yellowstone

### 3.1. Yellowstone breathes because it is alive

The "heavy breath" of the Yellowstone caldera with periodic (time scale of decades) uplift and subsidence with total amplitude of about 0.5 m (see references in [36]) was attributed to the exsolved water rich fluid at the solidification [48] or accumulation of two-phase CO$_2$- H$_2$O fluid in the upper 4 km subsurface layer [39]. Recent episode of the fast uplift in 2004-2007 was studied in detail [59] and interpreted as a result of the deep fluid and magma intrusion at the depths of about 7-12 km (horizontal size of expanding object was about 20 km). Pure fluid invasion was also anticipated near Norris Geyser Basin.

Probable connection with CO$_2$ fluxing as a cause of the slow Yellowstone chamber deformations is confirmed by the approximately fivefold increase in 1978 in the content of "magmatic carbon" in the annual rings of pines from Mud volcano thermal area [60] associated with earthquakes swarm and transition to the subsidence phase of the caldera surface. Certainly to reach shallow geothermal level CO$_2$ must pass through the mush zone. We will consider mechanical respond of this deep zone on CO$_2$ pumping that should have inertia larger than shallow reservoirs that can be in pace with several decades period of deformations. Similar mechanical disturbances of the surficial geothermal system (depth 2-3 km) caused by invasion of the fluid from intruded magma in Campi Flegri geothermal field (Italy) dissipate in the time scale of several years [61] by spreading excess fluid horizontally in Darcy's flow.

If deep CO$_2$ enters the mush zone by a mechanism more efficient than Darcy's filtration (e.g. with superheated rhyolites or through fracture zones), one can expect mush zone to expand. A rule of thumb estimate of the maximum value of the overpressure which can be reached in t years is $\Delta P = K_{ch}(\gamma q_{co2} S \Delta t / \rho_{fl} V_{ch})$, where $\gamma$ is correction factor accounting water extraction from the melt (here 1.2), $\Delta t$ is time interval of pumping (here 50 years), S – caldera area (2000 km$^2$, we use active area less than maximum one delineated by circular fault), $V_{ch}$ is magma chamber volume that can be taken for Yellowstone as 3000 km$^3$/0.3, K is bulk modulus of 20 GPa, $\rho_{fl}$ is density of fluid at P=2 kbar and T=850°C it is 600-500 kg/m$^3$. At $\Delta t$ = 50 yrs this order of magnitude estimate is 18 MPa. The coefficient $\gamma$ characterizes incoming fluid volume increase due to water extraction from the melt. As discussed above, its value depends on the actual volume of the melt that reacted with fluid. If interaction is efficient enough and the mass ratio of fluid/reacted melt is small this coefficient can be up to 4, otherwise it is in the range 1.1-1.2.

### 3.2. Viscoelastic model of flushing

It follows that at the used parameters values overpressure and associated deviatoric stresses can be high enough for mechanical failure and fluid release followed by the caldera subsidence and beginning of the new cycle. Considering the periodic expansion and contraction of the magma chamber in the time scale of decades, it is important to take into account the viscoelastic properties of heated surrounding rocks. As a first approximation of this problem, the formulation in the form of a viscoelastic spherical shell (approach used in [62]) can be analysed. With an increase of the relative shell thickness (the heated and damaged rocks aureole), this approximation at small times approaches the conventional equation for a spherical inclusion in an infinite elastic medium. Because of the spherical symmetry, this problem reduces to 1-D and has a simple analytical solution (see in the textbooks). The geometry of the shell is determined by the outer $R_1$ and the inner $R_0$ radii. On the external surface, pressure (radial stress) $p_1$ and $p_0$ is applied on the inner surface. It is easy to show that the stress distribution does not depend on the material parameters (elastic, viscous or viscoelastic) and depends only on the shell geometry:

$$s_2 = \frac{-P_1 R_1^3 (R^3 + R_0^3) + P_0 R_0^3 (R_1^3 + 2R^3)}{2R^3 (R_1^3 - R_0^3)}, \quad s_{rr} = -\frac{p_1 R_1^3 (R^3 - R_0^3) + p_0 R_0^3 (R_1^3 - R^3)}{R^3 (R_1^3 - R_0^3)} \quad (15)$$

here $s_2$ are angle components of stress tensor in spherical coordinates $s_{\varphi\varphi} = s_{\tau\tau}$. The radial displacement in an elastic solution is

$$u(R) = -P_1 \left(\frac{R^3}{3} + \frac{3K}{4\mu} R_0^3\right) R_1^3 + P_0 \left(\frac{R^3}{3} + \frac{K}{4\mu} R_1^3\right) R_0^3 / K(R_1^3 - R_0^3) R^2 \quad (16)$$

With a confining pressure $P_1$ equal to the internal $P_0$, the displacement field is reduces to a uniform displacement field corresponding to the volume deformation and all components of the stress tensor ($\sigma_{rr}, \sigma_{\phi\phi}, \sigma_{\theta\theta}$) becomes $-P$. In the expression for the tangential stress ($s_{rr}$), the first term, proportional to the confining pressure, is compressive, while the term proportional to the internal pressure is tensile so that $s_{rr} > 0$ for the thin shell and high overpressure. The solution for the radial distribution of displacements u(R) depends on the material parameters and can be used to calculate the deformation of the magma volume due to internal processes. The general viscoelastic solution is easily derived from an elastic one through its Laplace transform, obtained by replacing the shear modulus with the complex shear modulus and the boundary pressures on their Laplace transforms. With an instantaneous onset of the pressures $P_1$ and $P_0$ (dP=$P_0-P_1$), the expression for the transient displacements u($R_1$,t) at the outer boundary is a sum of

$$u_1(R_1, t) = dP \frac{\alpha^3 R_1 t}{4(1-\alpha^3)\eta}, \quad u_2(R_1, t) = dP \frac{R_1(1+\nu)}{3E\nu} \frac{(2-\nu)\alpha^3}{2(1-\alpha^3)}, \quad u_3(R_1, t) = -dP \frac{R_1(1-2\nu)^2 \alpha^3}{3E\nu(1-\alpha^3)} \exp\left(-\frac{3E\nu}{2(1+\nu)^2 \eta} t\right), \quad (17)$$

where the first term (eqn. 17) describes viscous deformations of the shell at the applied overpressure dP and two next terms describe viscoelastic deformations. Sum of $u_3$ and $u_2$ corresponds to some expansion due to exponential relaxation. Term $u_2$ corresponds to the residual displacement at the full relaxation. Elastic displacements $u_2+u_3$(t=0) (equivalent to the surface uplift at dP>0) are proportional to dP$R_1$/E. By setting $R_1$=3000 m, dP=30 MPa and E=30 GPa we get u=0.9 m initial deformations and 1.2 m residual uplift (at dP=const=30 MPa). Viscosity $\eta=10^{19}$ Pas can be considered as typical for the weak lower crust below brittle-ductile transition boundary. This boundary arises above active magma chambers. Viscosities $\eta \leq 10^{18}$ Pas delineate zone the closest thermal aureole. One can distinguish between this zone and magma mush only in a short time scale deformations. By setting effective viscosity of the shell of $\eta=10^{19}$ Pas we get viscous uplift (expansion) rate of 2 cm/y. At time interval 60 years viscous and elastic deformations equated in magnitude. Relaxation time scale at these parameters will be 44 years.

**3.3. Solution for magma pressure variation**

For the general dependence of the internal pressure $P_0+dP(t)$ on time, the expression for radial displacements is formulated with viscoelastic and viscous terms (retaining terms depending on the overpressure $dP(t)$) as:

$$u_{ve}(R_1,t) = \frac{R_1\alpha^3}{(1-\alpha^3)}\left(dP(t)\frac{3(1-\nu)}{2E} - \frac{(1-2\nu)^2}{2\eta(1+\nu)^2}\int_0^t \exp\left(-\frac{3E\nu(t-\xi)}{2(1+\nu)^2\eta}\right)dP(\xi)d\xi\right) \qquad (18a)$$

$$u_{ve}(R_0,t) = \frac{R_0\alpha^3}{(1-\alpha^3)}\left(dP(t)\frac{1+\nu+2\alpha^3(1-2\nu)}{2E\alpha^3} - \frac{(1-2\nu)^2}{2\eta(1+\nu)^2}\int_0^t \exp\left(-\frac{3E\nu(t-\xi)}{2(1+\nu)^2\eta}\right)dP(\xi)d\xi\right) \qquad (18b)$$

$$u_{visc}(R_1,t) = \frac{R_1\alpha^3}{4\eta(1-\alpha^3)}\int_0^t dP(\xi)d\xi, \quad u_{visc}(R_0,t) = \frac{R_0}{4\eta(1-\alpha^3)}\int_0^t dP(\xi)d\xi \qquad (18c)$$

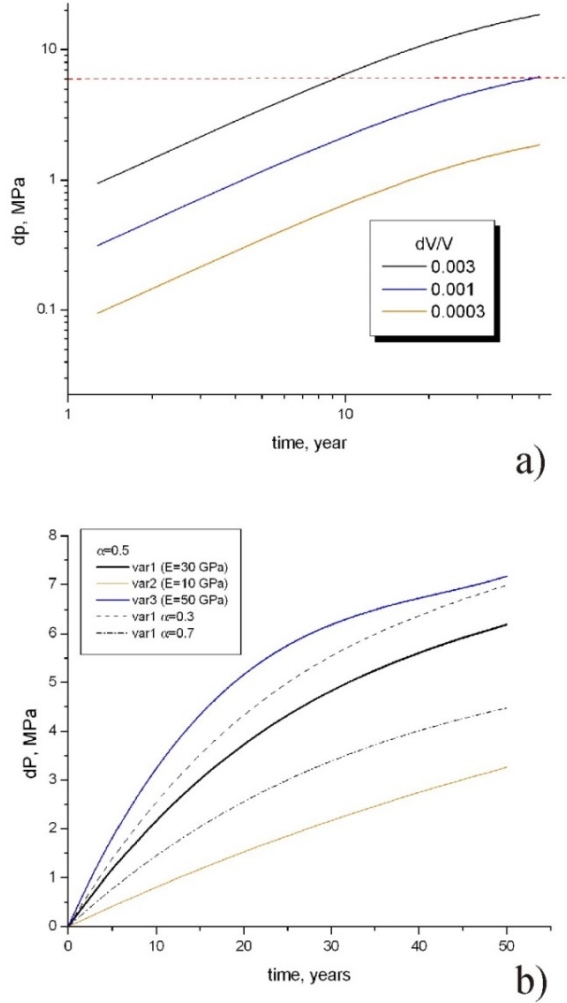

**Figure 8.** Calculated pressure increase dP at the infilling of magma chamber with $CO_2$ at P=2 kbar. Simplified model of a spherical viscoelastic shell filled with magma is used (see text). Different inflation rates are applied overpressure is calculated. In 50 years, the magma volume is increased by dV/V a) dV/V is in the range 0.0003-0.003 b) dV/V=0.001 and different values of Young's modulus, E, and relative thickness of the shell ($\alpha=R_1/R_0-1$) are used.

Relative volume increment (dV/V) in time t then can be expressed through displacement of the inner surface as dV/V=3($u_{ve}(R_0,t)+u_{visc}(R_0,t)$)/$R_0$ . Equations (18) are transformed into nondimentional form with pressure scale $P_{sc}$ (50 MPa) and time scale $t_0$ (10 years):

$$\frac{dV}{V} = \frac{3u(R_0)}{R_0} = \frac{3\alpha^3}{(1-\alpha^3)}\left(c_1 dp(\tau) - c2\int_0^\tau e^{c3(\tau-\xi)}dP(\xi)d\xi + \frac{3c4}{(1-\alpha^3)}\int_0^\tau dP(\xi)d\xi\right), \qquad (19)$$

where coefficients are:

$$c_1 = \frac{1+\nu+2\alpha^3(1-2\nu)P_{sc}}{2E\alpha^3}, c_2 = \frac{P_{sc}t_0(1-2\nu)^2}{2\eta(1+\nu)^2}, c_3 = \frac{3E\nu t_0}{2(1+\nu)^2\eta}, c_4 = \frac{P_{sc}t_0}{4\eta} \quad (20)$$

and can be expressed through two independent physical parameters namely, the Deborah number, De=$Et_0/\eta$, and nondimensional Young modulus, $E/P_{sc}$. A geometric factor $\alpha=R_1/R_0-1$ is another independent variable. Poisson's ratio, $\nu$, is in the range 0.2-0.25. We use the physical parameters values as in the examples considered above, i.e., Young's modulus E=30 GPa, viscosity $\eta=10^{19}$Pas, Poisson's ratio $\nu$=0.25, $\alpha$=0.5. To get the simplest solution we take a polynomial representation of the overpressure in nondimentional time:

$$dp(\tau) = \sum_{i=1}^{3} a_i \tau^i \quad (21)$$

Upon substituting $dp(\tau)$ in eqn.(19), it takes a closed form. Then, it becomes easy to find an approximation that produces close to linear in time volume growth at a constant fluid flux $dV/V(\tau)$ in the interval $0<\tau<5$ by minimizing $\int_0^\tau (dV(dp(\xi))/V_{ch} - \gamma q_{fl}\xi/V_{ch})^2 d\xi$. This is easy to do in Maple, and the order of magnitude of the mean integral square deviation for the third order polynomial is $10^{-12}$.

Relative magma chamber expansion dV/V at $\tau$=5 (t=50 years) is the controlling parameter depending on the multiplication coefficient $\gamma$ and $CO_2$ flux. Indeed, even the chamber volume $V_{ch}$ is not well constrained. We take $V_{ch}$ in the range (2 – 3) $10^4$ km$^3$, $\gamma$ = 1.2-4, S=1000-2000 km$^2$, $q_{CO_2}$ =40 kg/m$^2$/yr and get at $\tau$=5

$$dV/V = \frac{\gamma \tau q_{CO_2} S}{V_{ch}\rho_{fl}} = 0.5 \cdot 10^{-4} - 0.002 \quad (22)$$

With variation of $q_{CO2}$ in the range 0-100 kg/m$^2$/yr, the relative volume increase is within 0-0.005. We calculate the variation of overpressure with time for different parameters values (see Fig.8a). For a fixed set of mechanical parameters and different $CO_2$ fluxes (dV/V=(0.3- 3)$10^{-3}$) the overpressure is 1-11 MPa. Hydraulic fracturing with accumulated $CO_2$ starts at an overpressure that exceeds the tensile strength of rocks, which is realistically estimated to be around 6 MPa [63]. It means that in 10-50 years, $CO_2$ will be released into the upper crust and after uplift the caldera will subside. Indeed, mechanical parameters and relative shell thickness also influence the pressurization rate (see Fig.8b), and the given calculations can't be considered as an irrefutable proof for the discussed mechanism of "heavy breath". Nevertheless, results of these calculations show that for a certain range of parameters values the mechanism can correctly explain the observations. More field data is required to unambiguously distinguish between accumulation of residual fluid at the solidification [48], and $CO_2$ infilling. The first mode corresponds to the fast reduction of the magma volume while second to its growth or thermal equilibrium.

## 4. Acceleration of basaltic magma supply rate due to deglaciation

Large caldera-forming eruptions (LCFE) require a special eruptive mechanism, since normally overpressure in the chamber reaches zero when evacuating a very small fraction of the magma volume [63]. Dunson [63] has suggested that if an eruption occurs while the roof simultaneously subsides along outward dipping faults, then the overpressure can remain on a significant level. Indeed, it has been shown [1] that it is enough to have one outside dipping fault crossing the roof of the magmatic chamber to form a caldera with a trap-door configuration, which can later evolve into a full scale piston-like subsidence of the roof. At the early stage of an eruption, a strong overpressure is required to cause a critical failure of the roof of the magma chamber or the activation of a regional fault. Then, the integrity of the entire structure can be lost and pieces of the roof will literally float in the magma pool below. This will cause an abrupt drop in the magma pressure. When all the volume of the magma (or partial melt) is saturated with respect to the $H_2O$-$CO_2$ fluid (which is possible with a $CO_2$ fluxing), vesiculation will start throughout the magma volume and a reduced magma density will have a feedback effect on the pressure. In this scheme, the factor that creates a sharp increase in the rate of magma generation becomes a trigger capable of initiating LCFE.

### 4.1. Effect of deglaciation on the melting rate

Deglaciation is a factor, accelerating the rate of magma generation and strongly affecting the stress state of the crust. Glaciation in the Northern Hemisphere was initiated c.a. 3 Mys ago (e.g., [64]). Since then periodically large ice caps of up to several kilometers thickness were formed and melted in the high latitudes. Ice is similar to low density sedimentary rocks however, at the onset of the interglacial, ice melting rate exceeds normal denudation rates thus creating geologically unprecedentedly high unloading rates. Recently it was shown that deglaciation locally accelerates erosion rate that enhances net unloading effect [65]. In the areas of magma generation caused by decompressional melting in the ascending mantle flows (mantle plumes, compensatory flows at the delamination, back arc setting in subduction related mantle upwelling) fast decrease of the glacial load produces significant increase in the melting rate. Decompression melting in the ascending mantle flow occurs at a characteristic pressure decrease rate of $(3-5)10^{-4}$ MPa/yr at a flow velocity of 1.0–1.5 cm/yr. [66]. Thus, it appears that when the glacier melting time is about 1000 years, the associated decompression rate is 0.01 MPa/a. That is, the expected increase in melting rate is 20-30 times. In accordance with this estimate, the volumetric eruption rate in Iceland increased 20 times in the Holocene [67]. The effect of deglaciation on the Yellowstone giant magmatic chamber will be considered next.

### 4.2. Quaternary glaciations in Yellowstone

Geologic observations in the north-central United States (Iowa, Nebraska, Kansas and Missouri) [68] reveal that some of the tills that marked termination of major glaciation episodes were covered by ashes from the last Yellowstone eruptions (2.1, 1.2 and 0.64 Myr ago). Geographically studied region is are far away from Yellowstone but great advance of the ice cap from Canada recorded there probably was in phase with development of glaciation in the highlands in western United States including Yellowstone. Probably multiplicity of caldera forming eruptions at this location of the mantle plume contrasted with single major eruptions in other calderas of Snake River plain is related to the influence of deglaciation. The last young post LCT rhyolites eruptions of Central Plateau started in c.a. 260, 176, 124, 79 Kyr. [69]. All of them are well correlated with beginning of the global interglacials (see Fig. 9) after a maximum of 262, 182, 132,

93 Kyr. as recorded in the ice core of Vostok site of Antarctica ice shield [70]. Obviously, glacial state of the Yellowstone can differ from the global average reflected in the ice core. Delay of about 14 Kyr. for the last event can partially be linked with deviation from the integrated climate variation record. It is also unclear why not all global interglacials have paired rhyolitic eruptions episodes. Christiansen [71] presented arguments that some Central Plateau Member rhyolites were emplaced against glacial ice, that may imply that not all eruptions were initiated by deglaciation. These estimates are very preliminary and need extensive field studies to search glacial and postglacial signatures onsite.

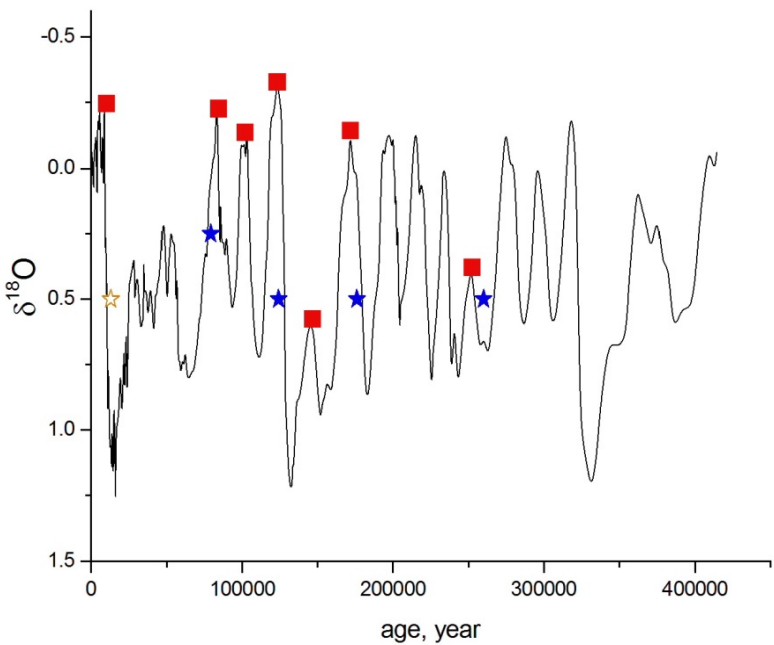

**Figure 9.** Isotope composition $\delta^{18}O$ of the oxygen from air in ice the core from the Vostok station, in East Antarctica [70] is plotted as global glaciation index, red squares are at the total ice mass minima (interglacials). Blue stars indicate time of the beginning of the recent volcanic activity in Yellowstone [69]. Semi-filled circle corresponds to the burst of the hydrothermal explosions in Holocene [72].

*4.2.1. Last Pleistocene glaciation in Yellowstone*

More is known about the last phase of Pleistocene glaciation in Yellowstone. Based on the fluid inclusions study [48] it has been concluded that Yellowstone plateau was covered with ice sheet with thickness 450-750 m. However, there were no eruptions that can be expected in connection with at least 10 fold increase of the generation rate of basaltic magma. Only hydrothermal explosions producing craters at most 2x3 km in size occurred in early Holocene [72]. First moment after ice load removal is characterized by the largest differential stresses at the depths close to the half-width of a glacier. In Holocene, strong earthquakes with magnitude larger than 6 were recognized even in the tectonically passive platforms [73]. Such tectonic events can produce new fractures changing the stress state of the hydrothermal reservoir and provide paths for the deep fluid, thus increasing heat transfer. Hydrothermal explosions are commonly explained by the pressure drop in the water dominated hydrothermal system with temperature close to boiling [74]. For example, abrupt decrease of the water table due to drainage of glacially dammed lakes [72] will induce deep boiling, reservoir expansion and mechanical disruption with boiling front propagating downwards. In their review Browne, Lawless [74] noted that accumulation of gaseous $CO_2$ near the surface may enlarge the effect of hydrothermal explosion in classic mode. Similar effects can be produced by the fast invasion of the $CO_2$ rich hot fluid into water dominated hydrothermal reservoir. A similar mechanism was proposed by Hedenquist and Henley [75], suggesting that Waiotapu (New Zealand) geothermal system was under stress when injected by magmatic gases that induced strong hydrothermal explosion (crater diameter 5 km, maximum depth of ejected clasts 350 m).

*4.2.1.1. Some geochemical evidences for $CO_2$ fluxing inYellowstone*

Reduced carbonic fluid is a good solvent not only for Pt but for Sr as well as other elements [76]. Thus, one can expect that $CO_2$ fluid passing through the mush zone would transfer LREE, Sr from the basalt magma to rhyolitic magma and extract Ba, Sr and some other elements from it, carrying them to the near-surface hydrothermal system.

Our suggestions are consistent with the geochemical data on hydrothermally altered rhyolites from Mary Bay (Yelowstone Lake) [72]. These altered rocks are enriched in Ba (up to 18 times) and especially Sr (up to 80 times) in comparison with intracaldera rhyolites, as expected if their transfer by deep $CO_2$ is significant. The Sr concentrations in the altered pyroclastics are much higher than in the extracaldera basalts [23]. More interestingly, these rocks (former rhyolites) have a high Ni content up to 1600 ppm. A close positive correlation between the content of Ni and Co, Ni and Cr, Ni and V was encountered. These observations suggest that some organometalic-compounds of these siderophilic elements were present in the fluid. A detailed hydrological study [77] revealed a clear outflow of Ba and Sr (extracted from the rhyolite mush) into the Yellowstone Lake with geothermal fluids.

*4.2.1.2. Current state of Yellowstone magma chamber*

In the view of the high uncertainties in the estimates of the heat and $CO_2$ fluxes and basaltic magma supply rate, results of the glacial probes which reflect the integrated state of the magma chamber under Yellowstone are of exceptional importance. Mechanical test of the last glaciation demonstrated that currently, ther magmatic system of Yellowstone is far from the impending catastrophic eruption. Impulse of the basaltic magmatism inevitably accompanying last deglaciation was dumped in the massive mush zone with low partial melt content. It is noteworthy that the time interval of rhyolite eruptions in the post LCT series [69] monotonously decreased with time, which can be interpreted as manifestation of the decay in the basaltic magma generation under the caldera. Moreover "periodic deglaciation forced evacuation" of magma from Yellowstone magma chamber might prevent it from a major eruption.

## 5. Conclusions

Since extra-large magma chambers require extra-large volumes of the added magma in a short time to erupt, eruptions have an extremely low frequency. Some large magma chambers, such as Yellowstone, are located in areas of not only exceptional magma generation rate, but also of very high $CO_2$ inflow which can significantly affect the mechanical stability of the chamber.

Deep $CO_2$ flux in the measured range alone can't prevent the cooling and solidification of the Yellowstone mush zone. In combination with the thermal energy of basaltic magma, a temporary increase in the flow of $CO_2$ can switch the thermal state from the thermal equilibrium to the mush melting.

Based on the analysis of melt inclusions, it is possible to estimate the current integrated mass flux of $CO_2$ to be at least 0.1-0.3 of the total melt mass in the mush zone of Yellowstone. The geochemical consequence of the long $CO_2$ fluxing can be averaging the magma composition and reducing of the content of Sr and Ba extracted into the geothermal system.

Simple modelling of deformations of viscoelastic rocks of the thermal aureole of Yellowstone magma chamber herein, shows that deep $CO_2$ fluxing in a reasonable range of flow rate and mechanical parameters can lead to slow uplift and subsidence cycles with a full amplitude of about 1 meter and a period of several decades.

The main triggering mechanisms of the eruption are spikes in the rate of generation of basaltic magma caused by internal processes in the mantle plume underlying Yellowstone. In the Quaternary, repeated glaciation in the Northern Hemisphere became another factor that periodically increased magma generation rate 10-30 times in early inetrglacials. The last Pleistocene glaciation in Yellowstone caused only hydrothermal and not magmatic eruptions in the Holocene. Since the previous interglacials were marked by voluminous post-LCT rhyolite eruptions, the result of Holocene glacial probing may imply that the current state of the Yellowstone magma system is in a stable or decaying thermal regime.